\documentclass[twocolumn,twoside,slac]{revtex4}
\usepackage{graphicx}
\usepackage{fancyhdr}
\begin{document}

\title{The Clarens Web Services Architecture}

%

\author{Conrad D. Steenberg}
\author{Eric Aslakson, Julian J. Bunn, Harvey B. Newman,
Michael Thomas, Frank van Lingen}
\affiliation{California Institute of Technology, Pasadena, CA, 91125, USA}

\begin{abstract}
Clarens is a Grid-enabled web service infrastructure implemented to augment 
the current batch-oriented Grid services computing model in the Compound Muon 
Solenoid (CMS) experiment of the LHC. Clarens servers leverage the Apache web 
server to provided a scalable framework for clients to communicate with 
services using the SOAP and XML-RPC protocols. This framework provides security,
session persistent storage, service discovery, and call routing to back-end
services. As an implementation policy Clarens uses widely implemented standards
wherever possible instead of inventing new standards.

This paper describes the basic architecture of Clarens, while a companion paper
describes clients and services that take advantage of this architecture. More 
information and documentation is also available at the Clarens web page at
{\tt http://clarens.sourceforge.net}.

\end{abstract}
\maketitle
\thispagestyle{fancy}
\section{Introduction\label{intro}}
The ascendance of Grid-like \cite{grid_phys_today} technologies has been all but
necessitated by the sheer volume of data produced in both science and commerce. 
In response to this increased uptake in High Energy Physics in particular, the 
traditional batch-oriented implementations \cite{globus_toolkit}\cite{sun_grid}
using home-grown protocols have started to adapt to an industry-wide move to 
standardized interfaces and protocols \cite{globus_ogsa}.

\begin{figure}[thb]
\begin{center}
\includegraphics[width=77mm, height=80mm, bb=20 239 534 728, clip=false]
 {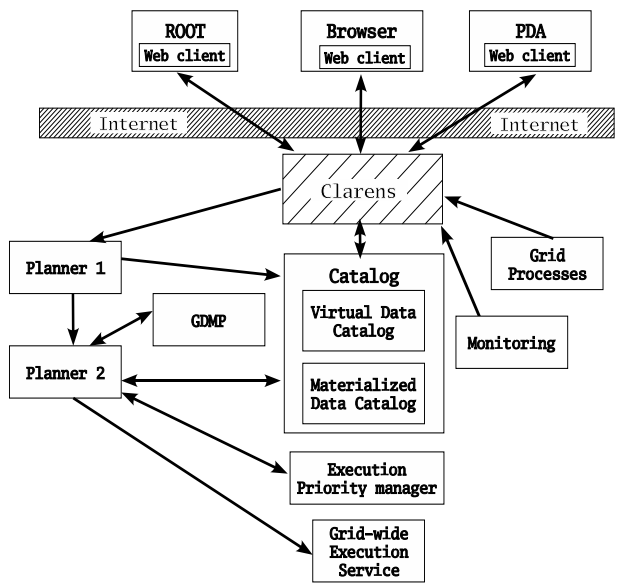}
\end{center}
\caption{CAIGEE architecture diagram.} \label{caigee_diagram}
\end{figure}

In this context the CAIGEE \cite{caigee} project was started to develop a specific
application of Grid technologies to the area of interactive analysis by end-user
physicists. A diagram showing Clarens as the interface between distributed clients
and Grid services is shown in Figure \ref{caigee_diagram}. 

In keeping with its public standards focus, clients can include anything from an
end-user analysis package like ROOT \cite{root}, web browsers or even web-enabled 
PDAs. Other Clarens servers may also act as clients. 

Back-end services planned or implemented include data movement, Grid-wide 
execution planning and scheduling, cluster job scheduling as well as metadata
catalogs.


\section{Infrastructure}

In order to save development time and improve scalability, The Clarens server 
is implemented as an extension to the Apache \cite{apache} multi-process web 
server using the {\tt mod\_python} extension in the Python byte-code compiled
language. Most CMS sites have these components already installed on processing 
cluster head nodes as part of the Redhat 7.3-based OS used. Clarens itself is
both architecture and platform-dependent by virtue of using Python as an 
implementation language.

\begin{figure}[thb]
\begin{center}
\includegraphics[width=54mm, height=57mm, bb=115 300 400 662, clip=false]{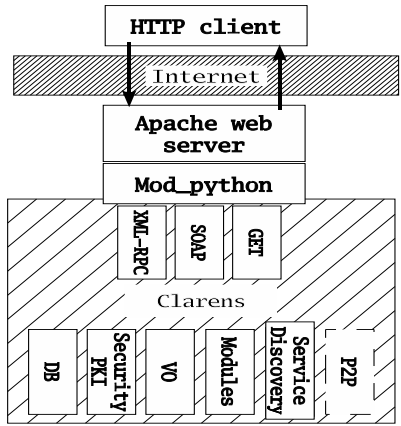}
\end{center}
\caption{Clarens architecture diagram.} \label{clarens_diagram}
\end{figure}

The Clarens architecture is depicted in Figure \ref{clarens_diagram} in the order
that requests are processed by the server. Firstly (top), the Apache server receives an 
HTTP 
POST or GET request from the client, and invokes Clarens based on the form of 
the URL specified by the client. Other URLs are handled as usual by the server 
according to its configuration. Secure Sockets Layer (SSL) encrypted connections
are handled transparently by the Apache server, with no special coding needed in 
Clarens itself to decrypt (encrypt) requests (responses). Encryption of network
traffic is optional, however, in cases where it is not required, without exposing
client or server credentials.

After the request has been processed, a response is sent back to the client, which
is usually encoded as an RPC response, but may also be in the form of binary 
data. GET requests returns a file or an XML-encoded error message to the client,
while XML-RPC \cite{xmlrpc} or SOAP\cite{soap}-encoded POST requests return a
similarly encoded response or error message.

\section{Authentication}
In keeping with the Grid tradition, Clarens uses a so-called Public Key 
Infrastructure (PKI) based authentication system that relies on X509 formatted
certificates issued by a Certification Authority (CA) along with asymmetric 
encryption using public and private keys. The authentication protocol of the 
server is implemented at the application level, thus 
eliminating the need for a custom security layer on the client side.

If an SSL-encrypted connection is used the client's certificate is provided to
the server as part of the connection negotiation stage. The Apache server passes
this information to the Clarens layer. This is the default for browser-based
clients which in general have a well developed client-side PKI security
infrastructure. In this case the authentication step in initiated by calling the
RPC method {\tt system.auth2()} with no arguments. A user session ID is requested by
the client by setting the {\tt clarens\_username} cookie value in the message 
header to the requested session ID, and setting the {\tt clarens\_password} 
cookie value to {\tt BROWSER}. Clarens responds by returning its own certificate
as well as it's part of the session ID encoded as an RPC response. In subsequent 
requests the client must set the {\tt clarens\_password} to this server session
ID. Browsers will automatically send these cookie values in subsequent requests.

In the case of an unencrypted connection, or a client not able to send it's
certificate as part of the connection negotiation phase, the session ID and 
client certificate must be sent using as the username and password in the the HTTP 
basic authentication header invoking the {\tt system.auth()} method. The server
responds with a list of (1) its certificate, (2) the server session ID encrypted
using the user's public key, and (3) the client session ID encrypted using the
server's private key. This ensures that only someone in possession of the client's
private key can discover the server session ID, and also that the server is in
possession of a private key matching the certificate sent as (1) above.
In subsequent requests the client should set the client session ID as username,
and server session ID as password in the HTTP basic authentication header.

Once this certificate and session ID exchange is completed, both the client and
server certificates can be verified against the publicly available CA 
certificate chain, knowing that the other party is in possession of a matching 
private key.

\section{Authorization}\label{sec_auth}
Authorization of clients to access server resources (invoking methods, accessing
files etc.) is done within the framework of a hierarchical Virtual Organization
(VO) with members uniquely identified by their Distinguished Names (DNs) issued 
by the CAs as part of all X509 certificates. 

\subsection{Virtual Organization}
\begin{figure}[tb]
\begin{center}
\includegraphics[width=75mm, bb=0 191 557 756, clip=false]{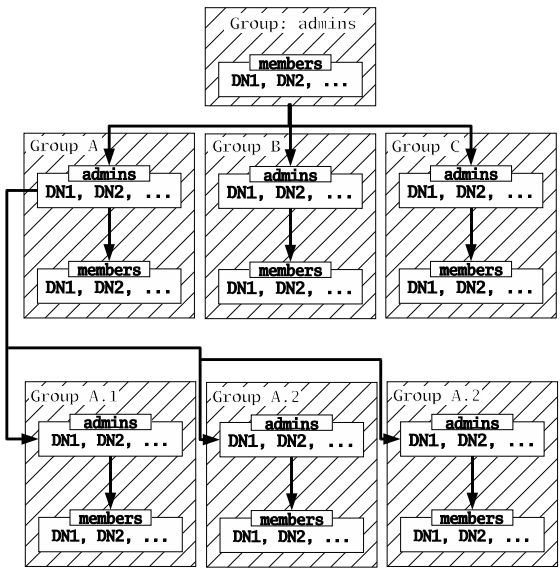}
\end{center}
\caption{Clarens virtual organization diagram.} \label{clarens_vo_diagram}
\end{figure}

Each Clarens server instance manages a tree-like VO structure, as shown in Figure
\ref{clarens_vo_diagram}, rooted in a list of administrators. This group, named 
{\tt admins}, is populated statically from values provided in the server configuration
file on each server restart. The list of group members is cached in a database
\cite{berkeley_db}, as is all VO information. The {\tt admins} group is
authorized to create and delete groups at all levels.

Each group consists of two lists of DNs for the group members and administrators
respectively. Group administrators are authorized to add and delete group members,
as well as groups at lower levels. The group structure is hierarchical because
group members of higher level groups are automatically members of lower level 
groups in the same branch.

The example in Figure \ref{clarens_vo_diagram} demonstrates the top-level groups
{\tt A}, {\tt B}, and {\tt C}, with second level groups {\tt A.1}, {\tt A.2}, and
{\tt A.3}.

A more concrete example might be to define groups {\tt CMS}, {\tt Atlas}, 
{\tt LHCb}, and {\tt Alice}, then for CMS, to define {\tt CMS.USA}, 
{\tt CMS.CERN}, {\tt CMS.UK}, {\tt CMS.Germany}. At the third level, one might
define {\tt CMS.USA.Caltech}, {\tt CMS.USA.UFL}, {\tt CMS.USA.FNAL}. Management
for the latter three groups may then be delegated to the institutes themselves,
thereby implementing a distributed trust model that has lower maintenance
overhead as well as being more representative of the real organizational
structure.

As a further optimization, the hierarchical information in the DNs may also be 
used to define membership, so that only the initial significant part of the DN
need to be specified. DNs are structured to include information on the country
(C), state/province (ST), locality/city (L), organization (O), organizational 
unit (OU), common name (CN), and e-mail address (Email). An example DN issued by
the DOE Science Grid CA is {\tt /O=doesciencegrid.org/OU=People/CN=John Smith 12345}
for individuals and {\tt /O=doesciencegrid.org/OU=Services/CN=host /www.mysite.edu}
for servers. To add all individuals to a particular group, only 
{\tt /O=doesciencegrid.org/OU=People} need to specified as a member DN.
\subsection{Access Control Lists}
Execution of RPCs (web service methods) as well as mapping of certificate DNs
to users on the server system is controlled by a set of hierarchical 
access control lists (ACLs) in a similar fashion to the VO structure described above,
and modeled after the access control ({\tt .htaccess}) files used by Apache.

Methods have a natural hierarchical structure modeled after Python's module
infrastructure. In fact all Clarens modules are also Python modules. Clarens 
places no arbitrary restrictions on the depth of this hierarchy, but a depth of
two or three levels is most common, e.g. {\tt module.method} or 
{\tt module.submodule.method}.

An ACL consist of an evaluation order specification (allow, deny or deny, allow)
followed by a list of DNs allowed, groups allowed, DNs denied and groups denied 
access. A DN or group granted access to a higher level method automatically has
access to a lower level method, unless specifically denied at the lower level, 
and is denied at the higher level unless allowed at a lower level. The ACL 
specification is therefore evaluated from the lowest applicable level to the
highest.

\begin{table}[tb]
\caption{Method ACL example.}\label{acl_table}
\begin{tabular}{|l|l|l|}
\hline
Object&Field&Value\\
\hline\hline
{\scriptsize\tt mod}&{\scriptsize\tt order}&{\scriptsize\tt deny, allow}\\
&{\scriptsize\tt allow DNs}& {\scriptsize\tt /O=doesg.org/OU=People/CN=John~Smith}\\
&&{\scriptsize\tt /O=doesg.org/OU=People/CN=Ng~Siong}\\
&{\scriptsize\tt allow groups}&{\scriptsize\tt CMS.USA}\\&&{\scriptsize\tt CMS.CERN}\\
&{\scriptsize\tt deny DNs}&{\scriptsize\tt /O=olduni/OU=physics/CN=Old Account}\\
&{\scriptsize\tt deny groups}&{\scriptsize\tt crackers}\\
\hline
{\scriptsize\tt mod.meth}&{\scriptsize\tt order}&{\scriptsize\tt deny, allow}\\
&{\scriptsize\tt allow DNs}&\\
&{\scriptsize\tt allow groups}&{\scriptsize\tt CMS.USA.Caltech}\\
&&{\scriptsize\tt CMS.USA.UFL}\\
&{\scriptsize\tt deny DNs}&{\scriptsize\tt /O=Caltech/OU=CACR/CN=Ed Peng}\\
&{\scriptsize\tt deny groups}&\\
\hline
\end{tabular}
\small
\end{table}

According to the example ACLs in Table \ref{acl_table}, when the method 
{\tt mod.meth} is invoked, the second ACL is applied, but when any other 
method in module {\tt mod} is invoked, the first ACL is applied.
\subsection{User Mapping}
In the traditional batch-oriented Grid architecture, being able to start 
long-running CPU intensive jobs as a certain system user is of crucial importance.
E.g. the Globus toolkit \cite{globus_toolkit} implements the concept of a so-called
{\em gridmap} file that maps system users to certificate DNs. It is implemented 
as a flat text file with two values per line, namely the DN and the system user.

Clarens similarly contains the notion of mapping DNs to user names. Instead of
a flat file, a structure similar to the ACLs described above is used, with the 
username taking the place of the method name, with one exception: the deny group
and deny individual fields are not used, since denying access to process creation
methods can be done using the method ACLs themselves.

At this point it should be clear that in both the VO, ACL, and user mappings
specification as little information as possible is stored in order to minimize 
search times for list memberships. Searching these lists are in the critical path
for the invocation of any method, though, and must be optimize as far as practical.
This is done by storing lists of DNs as strings in a structure called a {\em 
ternary tree} which is discussed in the appendix.
\subsection{Auditing}
All method invocations are logged with a time-stamp in the Apache server log files
to provide a record of client/server transactions.
\section{Modular Architecture}
Clarens provides a framework for extending its functionality via new modules
installed in subdirectories. Each subdirectory would appear as a new method,
root, e.g. a {\tt system} subdirectory would have its methods accessible as
{\tt system.{\em{}method}} \,etc. Modules can be implemented as either
interpreted Python bytecode, or as compiled C/C++ shared libraries if code 
execution speed is important, and are loaded on demand by the Python interpreter
in each Apache process, thereby providing crash-protection between different
processes, a major consideration for highly available servers.

Provision is also made for users on the server system to install their own
modules in subdirectories under their home directories. These modules are
accessible using the format {\tt $\sim$user.module.method}. 

Figure \ref{clarens_module} shows an example of the simplest module implementation
for a method {\tt echo}, that when placed in a file named {\tt echo/\_\_init\_\_.py}
returns its argument unchanged when called as {\tt echo.echo(argument)}.

\begin{figure}
\begin{verbatim}
from mod_python import apache
from clarens_util import 
  build_response, write_response

def echo(req,method_name,args):
    """Returns the method argument"""
    response=build_response(req,"echo",args)
    write_response(req,response)
    return apache.OK

methods_list={'echo':echo}
methods_sig={'echo':['string,string']}

\end{verbatim}
\caption{Clarens module example.}\label{clarens_module}
\end{figure}

Two important elements identifies this as a Clarens module: the 
{\tt methods\_list} and {\tt methods\_sig} structures identifying the list of 
methods and their signatures. Work is underway to use WSDL as method signatures
instead if the more limited XML-RPC signatures used in this example.
\section{Persistency}
Since the HTTP protocol does not {\em require}\footnote{ In the HTTP 1.1 
standard, persistent connections are the default for performance reasons, but 
the protocol is still inherently stateless, as opposed to e.g. the FTP 
protocol.} persistent connections, it is important that session information be 
stored persistently on the server side. This has the positive side-effect of 
allowing clients to survive server failures or restarts transparently without
having to re-authenticate themselves to the server in those cases.

The most important session information that is stored by the Clarens server is
the authentication information for each session. The Berkeley database 
\cite{berkeley_db} is used for this purpose.
\section{Scalability, failover, clustering}
Since Clarens is built upon commodity software components and standard operating
system services, it relies on these components to be set up to achieve these 
goals. Specifically, the Apache web, server, the filesystem, database and 
network components needs to be configured by the system administrator.

\section{Future developments}
Work is underway to extend Clarens from being an essentially client/server 
system to being a truly distributed system in a network of mutually aware
peers and superpeers that provide services. The most pressing need for large
scale Physics analysis is the need for truly distributed data
catalogs for a variety of Physics data and metadata, and of course a matching 
search capability.

A switch to a relational database for persistent data storage is planned to 
support more advanced data management than the Berkeley database's key/value 
mechanism can provide. Along with this change, file ACLs will also be 
implemented.

Werever practical Clarens aims to be compatible with the OGSA framework, with 
support for SOAP and WSDL being important first steps in that direction.

\section{Conclusion}
Clarens is a powerful, yet simple web services architecture with a strong 
emphasis on security for distributed virtual organizations. It draws upon a
rich base of commodity protocols and software components to provide a platform
for the deployment of analysis-oriented web services.

A companion paper describes services and clients that take advantage of this
platform \cite{clarens_apps}.

\begin{acknowledgments}

This work supported by Department of Energy contract DE-FC02-01ER25459, as
part of the Particle Physics DataGrid project \cite{ppdg}, and under
National Science Foundation Grant No. 0218937.

Any opinions, findings, and conclusions or recommendations expressed in this
material are those of the authors, and do not necessarily reflect the views
of the National Science Foundation.

Clarens development is hosted by SourceForge.net
\cite{sourceforge}.
\end{acknowledgments}

\section*{Appendix: Ternary trees}
As pointed out in Section \ref{sec_auth}, several lists must be searched for
every method invocation to enforce access control, establish group membership
and map certificate DNs to system user names.

\begin{figure}[hbt]
\begin{center}
\includegraphics[width=67mm, bb=0 0 390 460, clip=false]{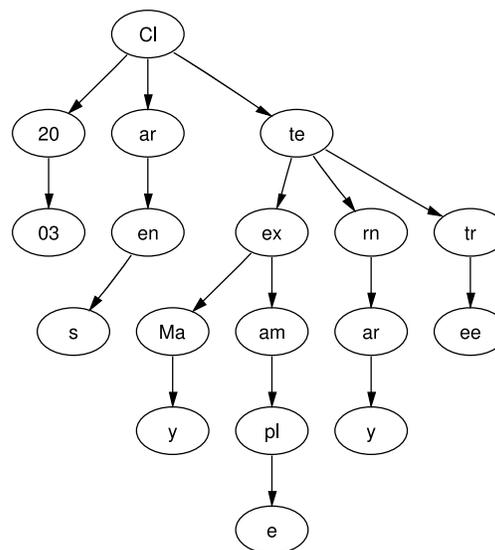}
\end{center}
\caption{Ternary tree graph for the words: {\em Clarens ternary tree example May 
2003}.} \label{ternary_tree}
\end{figure}

The keys for these lists (method, group, or user names) are kept in Python
dictionary objects, which are implemented as hash tables. Although it is
certainly possible to store the aforementioned lists in hash tables, these 
structures lack an important feature, namely the ability to search for
initial substrings. E.g. if we are presented with the query DN 
{\tt /O=myorg/OU=People/CN=John Smith}, but only the string 
{\tt /O=myorg/OU=People} is stored.

Initial substring searches are particularly suited to a data structure called a
{\em ternary tree} \cite{ternary_tree} that contains nodes for less than, equals
and greater than comparisons for string fragments. In contrast to the reference
implementation, the Clarens ternary trees contain two characters per node, 
reducing the storage requirements of the tree. 

When presented with the above DN, the tree can be traversed by comparing and
branching every two characters, and returning the number of characters matched
if a leaf node is reached, or a failure can be signaled if a leaf node is not 
reached. The returned number can be compared with the length of the query DN:
an exact string match corresponds to an equal number of matched characters,
while a lower number corresponds to a substring match.

The ternary tree structure is implemented as a Python extension module in C. 
Informal tests with 10,000 DNs on a 933 MHz PC produces roughly 
300,000 searches/second for the worst case where each DN is present in the
tree, i.e. the maximum number of branches taken per search.

\end{document}